\documentclass[authoryear,12pt,1p]{elsarticle}
\usepackage{amsthm,amssymb,amsmath}
\usepackage{setspace}

\newtheorem{lemma}{Lemma}
\usepackage{fancyvrb}
\DefineVerbatimEnvironment{Sinput}{Verbatim}{fontshape=sl}
\DefineVerbatimEnvironment{Soutput}{Verbatim}{}
\DefineVerbatimEnvironment{Scode}{Verbatim}{fontshape=sl}
\newenvironment{Schunk}{}{}

\usepackage[english]{babel}
\usepackage[T1]{fontenc}
\usepackage{url}
\usepackage[dvipsnames]{xcolor}
\usepackage{longtable}
\usepackage{rotating}

\usepackage{amsmath,amssymb}

\usepackage{algorithm}
\usepackage{algorithmic}
\makeatletter
\newcounter{algorithmbis}
\renewcommand{\thealgorithmbis}{\thesection.\arabic{algorithmbis}}
\def\algorithmbis{\@ifnextchar[{\@algorithmbisa}{\@algorithmbisb}}
\def\@algorithmbisa[#1]{%
  \refstepcounter{algorithmbis}
  \trivlist
  \leftmargin\z@
  \itemindent\z@
  \labelsep\z@
  \item[\parbox{\textwidth}{%
    \hrule
    \hrule
    \noindent\strut\textbf{Algorithm \thealgorithmbis} #1
    \hrule
  }]\hfil\vskip0em%
}
\def\@algorithmbisb{\@algorithmbisa[]}

\makeatother

\usepackage[all]{xypic}
\usepackage{enumerate}
\usepackage{bm}

\newcommand{\proglang}[1]{\texttt{#1}}
\newcommand{\pkg}[1]{\texttt{#1}}
\newcommand{\code}[1]{\texttt{#1}}
\newcommand{\R}{\ensuremath{\mathbb{R}}}

\newcommand{\abs}[1]{\ensuremath{\left\vert#1\right\vert}}
\newcommand{\var}{\ensuremath{\mathbb{V}\text{ar}}}

\newcommand{\E}{\ensuremath{\mathbb{E}}}
\newcommand{\pr}{\ensuremath{\mathbb{P}}}
\newcommand{\one}{\bm{1}}
\newcommand{\mvec}{\operatorname{vec}}
\newcommand{\tensor}[1]{#1^{\otimes 2}}

\newcommand{\VV}{\bm{\Omega}_\theta}
\newcommand{\mm}{\bm{\xi}_\theta}
\newcommand{\Ss}{\bm{\Sigma}_\theta}
\newcommand{\uu}{\bm{\mu}_\theta}

\newcommand{\M}{\bm{\mathcal{M}}}
\newcommand{\V}{\bm{\mathcal{V}}}      
\renewcommand{\phi}{\varphi}
\newcommand{\pdf}[1]{\ensuremath{\phi_{\bm{#1}}}}
\newcommand{\pdff}[2]{\ensuremath{\phi_{\bm{#1},\bm{#2}}}}
\newcommand{\cdf}[1]{\ensuremath{\Phi_{\bm{#1}}}}
\newcommand{\cdff}[2]{\ensuremath{\Phi_{\bm{#1},\bm{#2}}}}
\newcommand{\sminus}[1]{\ensuremath{{}^{\ominus}{\bm{#1}}}}
\newcommand{\splus}[1]{\ensuremath{{}^{\oplus}{\bm{#1}}}}
\newcommand{\nCDF}{\Phi_{\bm{\mu_\theta},\bm{\Sigma_\theta}}}

\newcommand{\diag}{\operatorname{diag}}
\newcommand{\independenT}[2]{
  \mathrel{\setbox0\hbox{$#1#2$}%
    \copy0\kern-\wd0\mkern4mu\box0}}

\usepackage{ifthen}
\newboolean{notesBoolean}
\setboolean{notesBoolean}{true} 
\newcommand{\mynote}[1]{%
  \ifthenelse{\boolean{notesBoolean}}{
    \begingroup
    \hbadness 20000
    \marginpar{\tiny\textsf{#1}}
    \endgroup
  }{}
}

\usepackage{xspace}
\newcommand{\hta}{\ensuremath{\text{5-HT}_{\text{2A}}}\xspace}

\newcommand{\BPp}{\ensuremath{{\text{BP}_{p}}\xspace}}

\usepackage{natbib}

\usepackage{graphicx}

  \title{A latent variable model with mixed binary and continuous
  response variables}

\begin{document}

\begin{frontmatter}


  \author[biostatku]{Klaus K\"{a}hler Holst\corref{cor1}}
  \ead{k.k.holst@biostat.ku.dk}
  \cortext[cor1]{Corresponding author.} 
  \author[biostatku]{Esben Budtz-J\o{}rgensen}
  \author[nru]{Gitte Moos Knudsen}
  \address[biostatku]{University of Copenhagen, Department of
    Biostatistics}
  \address[nru]{Neurobiology Research Unit and Center for Integrated
  Molecular Brain Imaging, Denmark}

  \begin{abstract}
    We propose a method for obtaining maximum likelihood estimates in
    a model with continuous and binary outcomes.
    Combinations of left and right censored observations are also
    naturally modeled in this framework.  The model and estimation
    procedure has been implemented in the \proglang{R} package
    \pkg{lava.tobit}.
    
    The method is demonstrated on brain imaging and personality data
    where measurement error on predictor variables is handled in a
    latent variable framework. A simulation study is conducted
    comparing the small sample properties of the MLE with a limited
    information estimator.
  \end{abstract}
\end{frontmatter}


\section{Introduction}
\label{sec:introduction}
Correlated binary data appears in a wide range of applications in
psychometrics and epidemiology, including questionnaire studies,
longitudinal studies, and cross-sectional studies with multivariate
measurements.  Depending on the application, different methods have
been introduced to model such data. Sometimes marginal effects are of
primary interest in epidemiology, in which
case inference based on generalized estimating equations (GEE)
\citep{liang_zeger_86} is preferable, and consistent estimates and
standard errors are obtained without the need for explicit modeling of
the variance structure of the outcomes.  In scale validation
studies (based on Item Response Theory), and for certain types of matched
analyses such as matched case-control studies, inference based on the
conditional maximum likelihood (CML) \citep{andersen1971} offers
several advantages due to its computational simplicity and the minimal
required assumptions on the distribution of the random effects.
However, when the assumptions for the CML are not fulfilled, or when
interest lies in conditional (subject specific) effects or quantification of the
actual variance components, other methods are needed.

A vast amount of literature and software has been written to deal with
this situation. The main challenge here is that the likelihood
function is given as an integral with respect to the random effects
and a numerical approximation to this intractable integral is
generally required. Approximate integrated likelihood methods have been
proposed via Adaptive Gaussian Quadrature (AGQ)
\citep{pinheirochao06}, and various variants of the
Expectation-Maximization-algorithm such as Monte Carlo EM (MCEM)
\citep{wei_tanner_90,mcculloch1997,song05:_multiv_probit_laten_variab_model}
and Stochastic Approximation EM (SAEM) \citep{Meza20091350}. Also
Bayesian methods have been popularized with the implementation of
general Gibbs-samplers \citep{gilks94:_bayes,plummer03:_jags}.

The dominating framework in applied statistics, however, seems to be
the AGQ framework, which is implemented in most widely used software
packages such as \proglang{stata GLLAMM} \citep{genmulsemrabehesketh},
\pkg{lme4} in \proglang{R} \citep{bates10}, and \proglang{SAS}
\pkg{PROC NLMIXED} \citep{sas}. In contrast most Monte Carlo methods
are based on more experimental implementations, where ad hoc decisions
on e.g. convergence has to be made. An attractive alternative is
available via the limited information estimator proposed by Muth\'en
\citep{muthen84limited} which generalizes classical structural equation models
to allow for inclusion of binary outcomes modeled via a Probit link.

Our model is based on the same principles as \citep{muthen84limited}. An
equivalent model formulation is a threshold model where the responses
are generated by underlying normal latent variables. This leads to
a generalization of tetrachoric correlations, where we allow
conditioning on both covariates and random effects.  Numerical
approximations is here limited to evaluations of orthant probabilities
of the multivariate normal distribution and with computational
complexity that is independent of the number of random effects in the
model. This is in contrast to the AGQ, where the complexity grows
exponentially in the number of latent variables (the curse of
dimensionality).

\section{The model}

\begin{figure}[htbp]
  {
    \centering
    \mbox{
      \includegraphics[height=7cm,keepaspectratio=true]{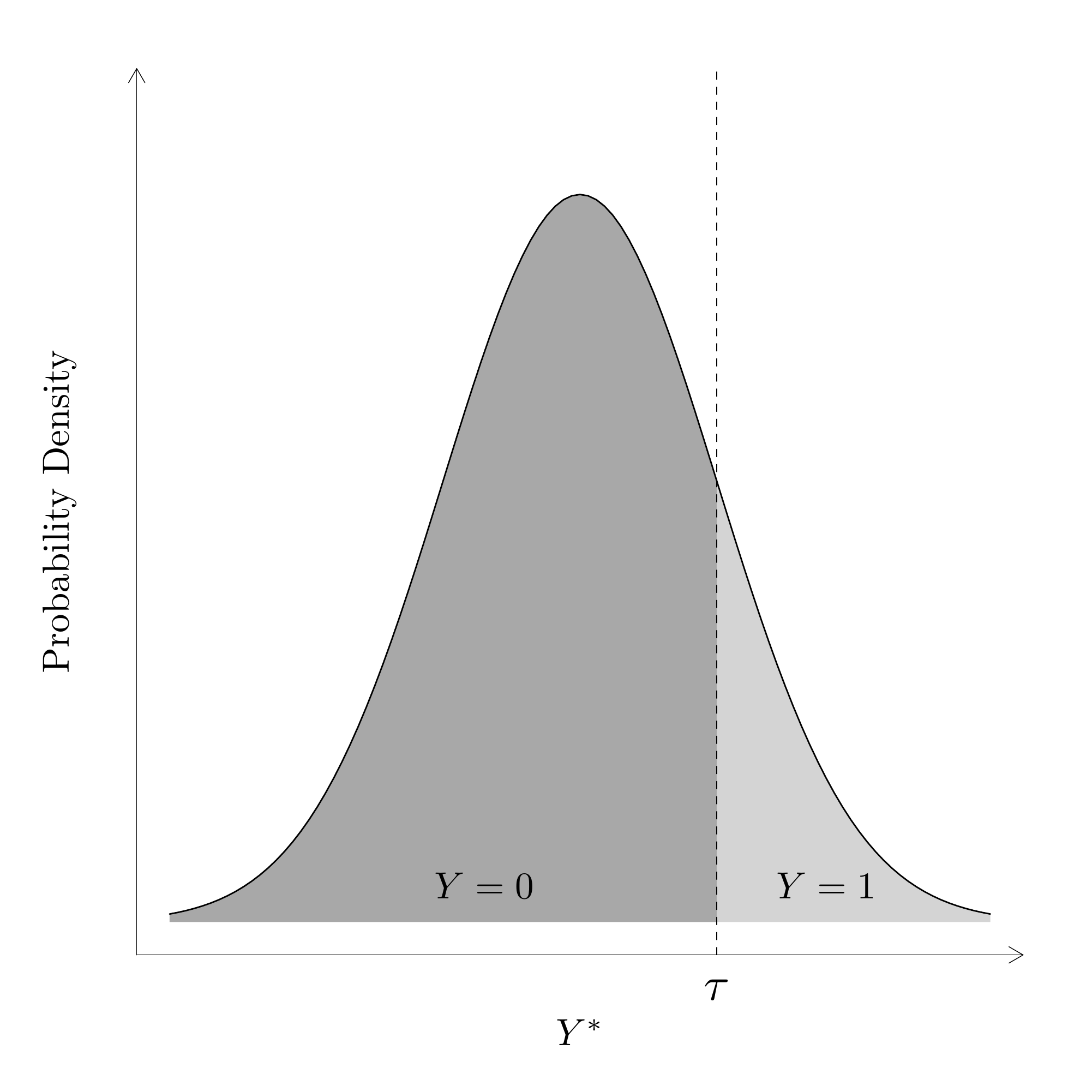}
      \includegraphics[height=7cm,keepaspectratio=true]{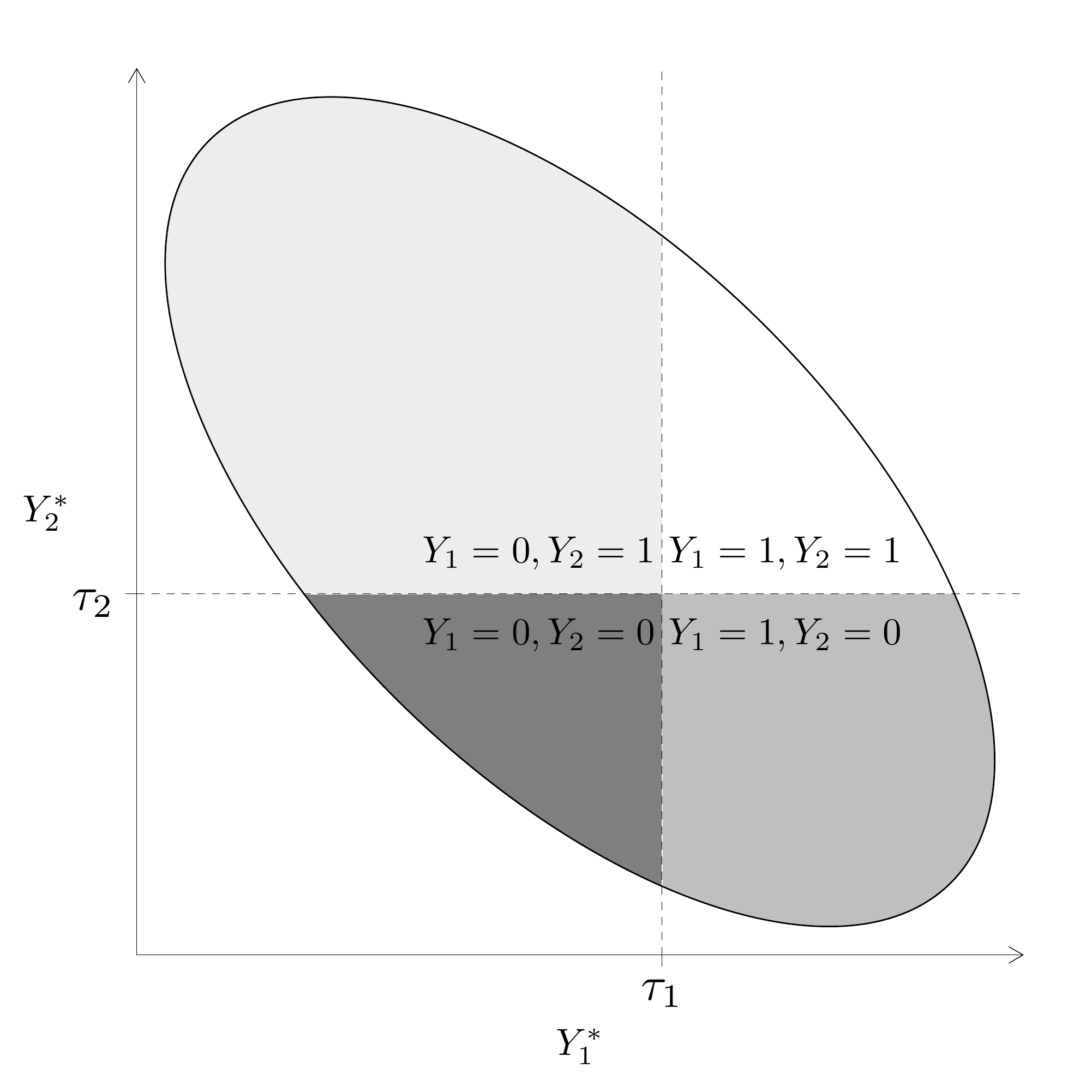}
    }
  }
  \caption{Univariate threshold model with probability of event for the
    observed binary variable $Y$ given by $\pr(Y=1) = \pr(Y^*>\tau)$. For the
    bivariate case $\pr(Y_1=1,Y_2=1) = \pr(Y_1^*>\tau_1, Y_2^*>\tau_2)$.}
  \label{fig:probit1}
\end{figure}

Let $Y_{ij}$ be a binary observation. The Probit model can be
formulated as a threshold model (see Figure \ref{fig:probit1}), where
we assume the existence of an underlying conditionally normally
distributed variable $Y_{ij}^*$ such that
\begin{align}\label{eq:thres1}
  Y_{ij} =
  \begin{cases}
    1,& Y_{ij}^*> 0 \\
    0,& Y_{ij}^*\leq 0.
  \end{cases}
\end{align}

We will let the underlying latent structure be described by
a general \emph{Linear Latent Variable Model} (LLVM), which can
be translated into a \emph{measurement part} describing the latent
responses, $Y_{ij}^*$:
\begin{align}
  \begin{split}\label{eq:summeas}
    Y_{ij}^* &= \nu_j + \sum_{k=1}^l\lambda_{jk}\eta_{ik} +
    \sum_{r=1}^q\kappa_{jr}X_{ir}
    + \sum_{k=1}^l  \delta_{jk}V_{ijk}\eta_{ik} \\
    &\qquad\qquad + \phi^{(j)}_{\bm{\theta}}(X_{i1},\ldots,X_{iq}) + \epsilon_{ij},
  \end{split}
\end{align}
given covariates $X_{ir}, V_{ijk}$, and 
a \emph{structural part} describing the random effects
\begin{align}
  \begin{split}\label{eq:sumstruct}
    \eta_{is} &= \alpha_{s} + \sum_{k}^l \beta_{sk}\eta_{ik} +
    \sum_{r=1}^q \gamma_{ir}X_{ir} + \sum_{k}^l
    \tau_{sk}W_{isk}\eta_{ik} \\
    &\qquad\qquad + \psi^{(s)}_{\bm{\theta}}(X_{i1},\ldots,X_{iq}) + \zeta_{ik},
  \end{split}
\end{align}
with \emph{individuals} $i=1,\ldots,n$ (between clusters
observations), \emph{measurements} $j=1,\ldots,p$ (within cluster
observations) and covariates $W_{isk}$ and random effects $\eta_{ik},
k=1,\ldots,l$.  For continuous observations we simply let
$Y_{ij}=Y_{ij}^*$.  Notice the interaction term
$\delta_{jr}V_{ijr}\eta_{ir}$ defines random slope components and
differentiate the LLVM from the a classic Structural Equation Model
formulation where the conditional variance of the responses given
covariates is assumed to be constant between individuals. Also the
functions $\smash{\phi_{\bm{\theta}}^{(j)}}$ and
$\smash{\psi_{\bm{\theta}}^{(s)}}$ allows for non-linear effects of
the covariates parameterized by $\bm{\theta}$. The random effects and
residual terms $\epsilon$ and $\zeta$ are assumed to follow normal
distributions, thus making the likelihood of the observed variables
available in explicit form. We will in details describe how to
evaluate the likelihood and score function of this model, and we will
compare the performance of our method to the limited information
estimator \citep{muthen84limited}.  The method is implemented in the
open-source \proglang{R} package \pkg{lava.tobit} \citep{lavatobit}.


\section{Estimation method}
\label{sec:method}
In the following subsections we will derive the likelihood and the
score function for a Gaussian latent variable model under
censoring. The Probit model for dichotomous outcomes follows
immediately as a special case.

\subsection{Right censored data}
\label{sec:right}
We first derive a multivariate extension of the Tobit regression model
for right censored normal distributed variables
\citep{tobin58}, and we later
show that this model immediately extends to the more general latent
variable framework.
          
Let $\bm{Y}\in\R^p$ be a multivariate censored response,
i.e. there exists an underlying variable $\bm{Y}^*$ and random
censoring times $\bm{C}=(C_1,\ldots,C_p)'$,
such that 
\begin{align}\label{eq:tobitright}
  Y_{j} =
  \begin{cases}
    Y_{j}^*, & Y_{j}^*< C_{j}\\
    C_{j}, & Y_j^*\geq C_{j}.
  \end{cases}
\end{align}

For a realization of $\bm{Y}$, we will partition the vector into
$\bm{y} = (\bm{y}_1',\bm{y}_0')'$ where $\bm{y_1}\in\R^m$ are fully
observed and $\bm{y_0}\in\R^k$ are actually censored components of
$\bm{Y_0}^*$.  We will assume that $\bm{Y}^*$ follows a normal
distribution $\mathcal{N}(\mm,\VV)$. In a common setup, we will also
wish to condition on a set of covariates, $(\bm{X},\bm{V},\bm{W})$,
but we will omit these here without loss of generality to avoid
unnecessarily complex notation.
                  
Let $\bm{\rho}=(\rho_1,\ldots,\rho_k)'$, and define the set 
\begin{align}
 \smash{\sminus{\rho} =
  \bigotimes_{i=1}^k ]-\infty,\rho_i]} \quad\text{and}\quad 
\smash{\splus{\rho} =
  \bigotimes_{i=1}^k [\rho_i,\infty[}.
\end{align}

Now the likelihood contribution (with a somewhat sloppy notation w.r.t
joint and conditional probability densities)
of a single observation is given by:
\begin{align}\label{eq:lik1}
  f_{\bm{\theta}}(\bm{y}_1,\bm{y}_0) = \int_{\splus{y_0}}
  \phi_{\bm{\theta}}(\bm{y}_1,\bm{u})\,d\bm{u}
  = \phi_{\bm{\theta}}(\bm{y}_1) \int_{\splus{y_0}} \phi_{\bm{\theta}}(\bm{u}\mid\bm{y}_1)\,d\bm{u}.
\end{align}
Obviously different individuals will typically have different patterns
of censoring, i.e. the partitioning of $\bm{Y}$ will be different, and
hence to calculate the log-likelihood of a complete data-set, summing
over the log-likelihood contributions of the different patterns of
censoring is necessary. 
          
In the following let $\pdff{\mu}{\Sigma}$ and
$\cdff{\mu}{\Sigma}$ denote the density and CDF, respectively, of the multivariate
normal distribution with mean $\bm{\mu}$ and variance $\bm{\Sigma}$
(with $\pdff{0}{\Sigma}=\pdf{\Sigma}$ and $\cdff{0}{\Sigma}=\cdf{\Sigma}$.).

In the case of no censoring the score is given by
\begin{align}
  \begin{split}\label{eq:S}
    \mathcal{S}(\theta;\bm{y}) &= -\frac{1}{2}\Big\{
    \left(\frac{\partial\mvec\VV}{\partial\bm{\theta}'}\right)'\Big(\mvec(\VV^{-1})
    \\
    &\qquad-
      \mvec\left[\VV^{-1}\tensor{(\bm{y}-\mm)}\VV^{-1}\right]\Big)
    \\
    &\qquad+
    2\left(\frac{\partial\mvec{\mm}}{\partial\bm{\theta}'}\right)'\VV^{-1}(\bm{y}-\mm)
    \Big\}.
  \end{split}
\end{align}

In the censored case, the derivative of the first term in the
log-likelihood corresponding to $\bm{Y}_1$
\begin{align}
  S_1(\bm{\theta};\bm{y}) = \frac{\partial}{\partial\bm{\theta}}\log\phi_{\bm{\theta}}(\bm{y}_1)
\end{align}
is easily obtained
from (\ref{eq:S}) by extracting the relevant sub-matrices of $\mm$ and $\VV$
and their derivatives, and it follows that
\begin{align}\label{eq:Scens}
  \mathcal{S}(\bm{\theta};\bm{y}) &= S_1(\bm{\theta};\bm{y}_1) + 
  \frac{\partial}{\partial\bm{\theta}'} \log
  \cdff{-\uu}{\Ss}(-\bm{y}_0) \\ &= S_1(\bm{\theta};\bm{y}_1) + 
  \frac{1}{\cdff{-\uu}{\Ss}(-\bm{y}_0)}\frac{\partial}{\partial\bm{\theta}'} 
  \cdff{-\uu}{\Ss}(-\bm{y}_0),
\end{align}
where $\uu$ and $\Ss$ are the mean and
variance of the conditional distribution of $\bm{Y}_0^*$ given
$\bm{Y}_1=\bm{y}_1$. 

We partition the 
mean and variance according to
the $(\bm{Y}_1,\bm{Y}_0^*)$:
\begin{align}
  \mm = 
  \begin{pmatrix}
    \mm^{(1)} \\
    \mm^{(0)}
  \end{pmatrix}
  \quad \text{ and } \quad
  \VV =
  \begin{pmatrix}
    \VV^{(1)} &     \VV^{(01)} \\
    \VV^{(10)} &     \VV^{(0)}
  \end{pmatrix},
\end{align}
hence
\begin{gather}
  \uu = \uu(\bm{y}_1) = \mm^{(0)} +
  \VV^{(01)}{\VV^{(1)}}{}^{-1}(\bm{y}_1 - \mm^{(1)}), \\
  \Ss = \VV^{(0)} - \VV^{(01)}{\VV^{(1)}}{}^{-1}\VV^{(10)},
\end{gather}
and it follows that
\begin{align}
  \begin{split}\label{eq:dcondmu}
    \frac{\partial\mvec\uu}{\partial\bm{\theta}'} &=
    \frac{\partial\mvec\mm^{(0)}}{\partial\bm{\theta}'}
    + \left\{(\bm{y}_1-\mm^{(1)})'\VV^{(1)}{}^{-1}\otimes\bm{1}_k\right\}\frac{\partial\mvec\VV^{(01)}}{\partial\bm{\theta}'} \\
    &\quad -
    \left\{(\bm{y}_1-\mm^{(1)})'\VV^{(1)}{}^{-1}\otimes\VV^{(10)}\VV^{(1)}{}^{-1}\right\}\frac{\partial\mvec\VV^{(1)}}{\partial\bm{\theta}'} \\
    &\quad -
    \VV^{(10)}\VV^{(1)}{}^{-1}\frac{\partial\mvec\mm^{(1)}}{\partial\bm{\theta}'}.
  \end{split}
\end{align}
\begin{align}
  \begin{split}\label{eq:dcondS}
    \frac{\partial\mvec\Ss}{\partial\bm{\theta}'} &=
    \frac{\partial\mvec\VV^{(0)}}{\partial\bm{\theta}'} -
    (\VV^{(01)}{\VV^{(1)}}{}^{-1}\otimes\bm{1}_k)\frac{\partial\mvec\VV^{(01)}}{\partial\bm{\theta}'}
    \\
    &\quad +
    (\VV^{(01)}{\VV^{(1)}}{}^{-1}\otimes\VV^{(01)}{\VV^{(1)}}{}^{-1})
    \frac{\partial\mvec\VV^{(1)}}{\partial\bm{\theta}'}
    \\
    &\quad - (\bm{1}_m\otimes\VV^{(01)}{\VV^{(1)}}{}^{-1})
    \frac{\partial\mvec\VV^{(10)}}{\partial\bm{\theta}'},
  \end{split}
\end{align}
                   
To calculate the score, (\ref{eq:Scens}), the derivative of the CDF
with respect to $\bm{\theta}$
is needed, and the result is given in Lemma \ref{lem:2} below.
We will need to define the operators
$\V_{\bm{\mu},\bm{\Sigma}}\colon\R^k\to\R^{k\times k}$ and
$\M_{\bm{\mu},\bm{\Sigma}}\colon\R^k\to\R^k$ as
\begin{gather}\label{eq:VM}
  [\M_{\bm{\mu},\bm{\Sigma}}(\bm{y})]_{i} =
  \int_{\sminus{y}} \pdff{\mu}{\Sigma}(\bm{x})(x_i-\mu_i)\,d\bm{x}, \\
  [\V_{\bm{\mu},\bm{\Sigma}}(\bm{y})]_{ij} =
  \int_{\sminus{y}} \pdff{\mu}{\Sigma}(\bm{x})(x_i-\mu_i)(x_j-\mu_j)\,d\bm{x}.
\end{gather}
Evaluating these quantities are obviously closely related to
calculating the moments of a truncated normal distribution.  We assume
without loss of generality that $\bm{\Sigma}$ is a correlation
matrix, and look at a standardized normal distribution
$\bm{W}\sim\mathcal{N}(\bm{0},\bm{\Sigma})$, and let $\bm{T}$ be the
right-truncated version of $\bm{W}$, such that $\bm{T}$ is equal to
$\bm{W}$ on the region $\sminus{y}$. From \citep{tallis1961} the
moment generating function of $\bm{T}$ is
\begin{align}
  m(\bm{t}) = \E(\exp^{\bm{t}'\bm{T}}) =
  \alpha^{-1}\exp(\tfrac{1}{2}\bm{t}'\bm{\Sigma t})\cdf{\Sigma}(\bm{y}-\bm{\Sigma
    t}),
\end{align}
with normalizing constant $\alpha=\cdf{\Sigma}(\bm{y})$, and hence the
first and second moments of $\bm{T}$ are
\begin{gather}
  \E(\bm{T}) = -\alpha^{-1}\bm{\Sigma}D\cdf{\Sigma}(\bm{y}) \\
  \E(\bm{T}\bm{T}') = \bm{\Sigma} + \alpha^{-1}\bm{\Sigma}H\cdf{\Sigma}(\bm{y})\bm{\Sigma},
\end{gather}
with $D$ and $H$ denoting the gradient and hessian operator.  To
calculate these quantities, we can exploit that for the stochastic
variables $(U,V)$ partial derivatives with respect to the first
components of the corresponding CDF, $F$, is simply the product of the
marginal probability density function of $U$ times the CDF of the
conditional distribution of $V$ given $U$, $\partial_u F(u,v) =
f(u)F(v|u)$.  As the conditional distributions of a normal
distribution is available in analytical form, the gradient and hessian
of $\cdf{\Sigma}$ can explicitly be calculated as functions of the CDF
and pdf only. Details can be found in \citep{tallis1961} and
\citep[supplementary]{vaida09:_fast_implem_for_normal_mixed}. Further,
as fast and precise approximations are available for the evaluation of
the multivariate normal CDF
\citep{genz92:_numer_comput_of_multiv_normal_probab,mvtnorm},
the evaluation of the integrals (\ref{eq:VM}) is an achievable task.
\begin{lemma}\label{lem:1}
  Assume $\Ss$ has full rank and let $\bm{R}_{\bm{\theta}} =
  \bm{\Lambda}_{\bm{\theta}}^{-1}\Ss\bm{\Lambda}_{\bm{\theta}}^{-1}$ be the corresponding correlation matrix
  where $\bm{\Lambda}_{\bm{\theta}}$ is the diagonal-matrix with $\diag(\bm{\Lambda}_{\bm{\theta}}) =
  \diag(\Ss)^{1/2}$. Then
  \begin{align*}
    \M_{\uu,\Ss}(\bm{y}) &=
    -\bm{\Lambda}_{\bm{\theta}}\bm{R}_{\bm{\theta}}D
    \cdf{R}\left(\bm{\Lambda}_{\bm{\theta}}^{-1}(\bm{y}-\uu)\right) \\
    \V_{\uu,\Ss}(\bm{y}) &= \cdff{\mu}{\Sigma}(\bm{y})\Ss \\ &\qquad+
    \bm{\Lambda}_{\bm{\theta}}\bm{R}_{\bm{\theta}}H
    \cdf{R}\left(\bm{\Lambda}_{\bm{\theta}}^{-1}(\bm{y}-\uu)\right)
    \bm{R}_{\bm{\theta}}\bm{\Lambda}_{\bm{\theta}}.
  \end{align*}
  \begin{proof}
    Follows immediately by substitution $\smash{\bm{v} = \psi(\bm{u}) =
      \bm{\Lambda}_{\bm{\theta}}^{-1}(\bm{u}-\uu)}$, noting that $\phi(\sminus{y})$ is
    scaled and translated but not rotated.
  \end{proof}
\end{lemma}
          
\begin{lemma}\label{lem:2}
  \begin{align*}
    \frac{\partial}{\partial\bm{\theta}}\nCDF(\bm{y})
    &= 
    \frac{1}{2}\left(\frac{\partial\mvec\Ss}{\partial\bm{\theta}'}\right)'\Big[
    -\mvec(\Ss^{-1})\nCDF(\bm{y}) + \\
      &\qquad (\Ss^{-1}\otimes\Ss^{-1})\mvec\{\V_{\uu,\Ss}(\bm{y})\}
      \Big] + \\
    &\qquad
    \left(\frac{\partial\mvec\uu}{\partial\bm{\theta}'}\right)'\Ss^{-1}
    \mvec\{\M_{\uu,\Ss}(\bm{y})\}
  \end{align*}

  \begin{proof}
    By the fundamental theorem of calculus 
    \begin{align}
      \begin{split}
      \frac{\partial}{\partial\bm{\theta}}\nCDF(\bm{y})
      &=
      \int_{\sminus{y}} \frac{\partial}{\partial\bm{\theta}}
      \phi_{\bm{\mu_\theta},\bm{\Sigma_\theta}}(\bm{u})\,d\bm{u} \\
        &=
        \int_{\sminus{y}}\frac{1}{(2\pi)^{k/2}}\frac{1}{\abs{\Ss}^{1/2}}\frac{-1}{2}
        \left(\frac{\partial\mvec\Ss}{\partial\bm{\theta}'}\right)' \\
        &\qquad 
        \times\mvec(\Ss^{-1})  \\
        &\qquad
        \times\exp\left(-\tfrac{1}{2}(\bm{u}-\uu)'\Ss^{-1}(\bm{u}-\uu)\right)\,d\bm{u}
        \\
        &+
        \int_{\sminus{y}}\frac{1}{(2\pi)^{k/2}}\frac{1}{\abs{\Ss}^{1/2}} \\
        &\qquad \times\exp\left(-\tfrac{1}{2}(\bm{u}-\uu)'\Ss^{-1}(\bm{u}-\uu)\right)
        \\
        &\qquad \times\Big\{
        \frac{1}{2}\left(\frac{\partial\mvec\Ss}{\partial\bm{\theta}'}\right)'
        \\
        &\qquad\times
        \mvec(\Ss^{-1}\tensor{(\bm{u}-\uu)}\Ss^{-1}) \\
        &\qquad +
        \left(\frac{\partial\mvec\uu}{\partial\bm{\theta}'}\right)'\Ss^{-1}(\bm{u}-\uu)
        \Big\}\,d\bm{u},
      \end{split}
    \end{align}
    where we have used that
    \begin{align}
      \frac{\partial}{\partial\bm{\theta}}\abs{\Ss}^{-\tfrac{1}{2}}
      = -\frac{1}{2}\abs{\Ss}^{-\tfrac{1}{2}}
      \left(\frac{\partial\mvec\Ss}{\partial\bm{\theta}'}\right)'
      \mvec\left(\Ss^{-1}\right)
    \end{align}
    and
    \begin{align}
      \begin{split}
        &-\frac{\partial}{\partial\bm{\theta}}\left((\bm{u}-\uu)'\Ss^{-1}(\bm{u}-\uu)\right)
        = \\
        &\qquad\left(\frac{\partial\mvec\Ss}{\partial\bm{\theta}'}\right)'
        \mvec(\Ss^{-1}\tensor{(\bm{u}-\uu)}\Ss^{-1})
        \\
        &\qquad +
        2\left(\frac{\partial\mvec\uu}{\partial\bm{\theta}'}\right)'\Ss^{-1}(\bm{u}-\uu)
      \end{split}
    \end{align}
    derived by using standard matrix derivative results
    \citep{MR940471}.
    
  \end{proof}
\end{lemma}

\subsection{Combinations of left and right censored data}
\label{sec:leftandright}
The extension to include left censored variables follows almost
immediately from the right censored case.
As before we let $\bm{Y}\in\R^p$ be a multivariate censored response,
with underlying normal response $\bm{Y}^*\sim\mathcal{N}(\mm,\VV)$,
defined from \emph{right censoring time points} $\bm{C}=(C_1,\ldots,C_p)'$
and \emph{left censoring time points}
$\widetilde{\bm{C}}=(\widetilde{C}_1,\ldots,\widetilde{C}_p)'$, such that
\begin{align}
  Y_{j} =
  \begin{cases}
    Y_{j}^*, & \widetilde{C}_{j}<Y_{j}^*< C_{j}\\
    C_{j}, & Y_i^*\geq C_{j} \\
    \widetilde{C}_{j}, & Y_j^*\leq \widetilde{C}_{j}.
  \end{cases}
\end{align}
Again we will partition an observation into 
$\bm{y} = (\bm{y}_1',\bm{y}_0',\widetilde{\bm{y}}_0')'$, where
$\bm{y}_0\in\R^k$ are actual right censored and
$\widetilde{\bm{y}}_0\in\R^l$ are actual left censored, and we exploit
that the conditional distribution of the censored indices in $\bm{Y}$ given
the fully observed indices is $\mathcal{N}(\uu,\Ss)$.

We define the diagonal matrix
\begin{align}\label{eq:Lmat}
  \bm{L} = \one_k \oplus (-\one_l)
\end{align}
such that
\begin{align}
  \begin{split}
    &\int_{\sminus{\bm{y}_0}\times\splus{\widetilde{\bm{y}}_0}}
    \pdff{\uu}{\Ss} (\bm{u}_1,\bm{u}_2) \,d\bm{u}_1 d\bm{u}_2
    = \\
    &\qquad
    \int_{\sminus{\bm{y}_0}\times\sminus{-\widetilde{\bm{y}}_0}}
    \pdff{\bm{L}\uu}{\bm{L}\Ss\bm{L}}(\bm{u}_1,\bm{u}_2) \,d\bm{u}_1
    d\bm{u}_2,
  \end{split}
\end{align}
hence the log-likelihood is given by
\begin{align}
  \log \pdf{\bm{\theta}}(\bm{y}_1) + \log\cdff{\bm{L}\uu}{\bm{L}\Ss\bm{L}}
  \begin{pmatrix}
    \bm{y}_0\\
    -\widetilde{\bm{y}}_0 
  \end{pmatrix}
\end{align}
which is identified as the likelihood of right-censored data, and the
expression for the score function follows from the previous section,
noting that when applying Lemma \ref{lem:2}, we use that
\begin{gather}
  \frac{\partial\mvec \bm{L}\Ss\bm{L}}{\partial\bm{\theta}'} =
  \bm{L}^{\otimes 2}\frac{\partial\mvec\Ss}{\partial\bm{\theta}'}, \\
  \frac{\partial\mvec \bm{L}\uu}{\partial\bm{\theta}'} =
  \bm{L}\frac{\partial\mvec \uu}{\partial\bm{\theta}'}.
\end{gather}

We have not explicitly covered the situation with random effects in the
model (as in (\ref{eq:summeas})) but the observed data likelihood
follows immediately from the marginalization property of the normal
distribution.  Assume that $\bm{y} = (\bm{y}_1,\bm{y}_0)$ are fully
observed resp. right censored observation from an underlying normal
model $f_{\bm{\theta}}(\bm{Y},\bm{\eta})$, with random effects
$\bm{\eta}$, then by Fubini's theorem
\begin{align}
  \begin{split}
    \mathcal{L}({\bm{\theta}};\bm{y}) &= \int_{\R^k}\int_{\splus{y_0}}
    f_{\bm{\theta}}(\bm{y_1},\bm{u},\bm{\eta})\,d{\bm{u}}\,d\bm{\eta} \\
    &= \int_{\splus{y_0}}\int_{\R^k}
    f_{\bm{\theta}}(\bm{y_1},\bm{u},\bm{\eta})\,d{\eta}\,d\bm{u} \\
    &= \int_{\splus{y_0}}
    f_{\bm{\theta}}(\bm{y_1},\bm{u})\,d{\bm{u}} \\
    &= f_{\bm{\theta}}(\bm{y_1})\int_{\splus{y_0}}
    f_{\bm{\theta}}(\bm{u}\mid\bm{y_1})\,d\bm{u}.
  \end{split}
\end{align}
The last expression is identical to (\ref{eq:lik1}) for which we
have derived the score.

\subsection{Binary responses - the Probit model}
\label{sec:binary}          
The multivariate Probit is trivially handled by noting that all
observations are either left censored at $0$ (failure) or right
censored at $0$ (event). The probability of observing
$\bm{y} = (y_1,\ldots,y_p) \in\{0,1\}^p$ can therefore be expressed as the orthant
probability defined by the model-specific mean and variance
\begin{align}
  \pr(\bm{Y} = \bm{y}) = \int_{\otimes_{i=1}^p (\bm{1}_{\{y_i=1\}}-\bm{1}_{\{y_i=0\}})]0;\infty[} \phi_{\bm{\mu},\bm{\Sigma}}(\bm{x})\,d\bm{x}.
\end{align}
In terms of (\ref{eq:summeas}) this leads to the model
\begin{align}
  \begin{split}
    &\pr(Y_{ij}=0\mid \bm{X}_i,\bm{V}_i,\bm{\eta}_i) = \Phi\Big(\nu_j
    + \sum_{k=1}^l\lambda_{jk}\eta_{ik} +
    \sum_{r=1}^q\kappa_{jr}X_{ir} \\ &\qquad\qquad + \sum_{k=1}^l
    \delta_{jk}V_{ijk}\eta_{ik} +
    \phi^{(j)}_{\bm{\theta}}(X_{i1},\ldots,X_{ir})\Big),
  \end{split}
\end{align}
where $\Phi$ is the CDF of the standard normal distribution. Hence, in
this parameterization we are assuming that the residual terms of
$Y_{ij}^*$, $\epsilon_{ij}$, $j=1,\ldots,p$ follow a standard normal
distribution (not necessarily independent), and the thresholds are
fixed at zero as in (\ref{eq:thres1}). Other parameterizations are
possible, e.g. by constraining $\var(Y_{ij}^*)=1$, and letting the
intercepts of the model be zero.

\subsection{Implementation}
\label{sec:implementation}
The maximum likelihood estimates can be obtained via a
$BH^3$-algorithm \citep{bhhh} or one of several generic optimization
routines available in \texttt{R} \citep{team10:_r} (e.g the function
\code{nlminb}). Letting $\mathcal{S}_i(\bm{\theta})$ denote the
score contribution of the $i$th individual, we can obtain estimates of
the standard errors of the estimates, $\widehat{\bm{\theta}}$, via the
information matrix calculated as the outer product of the score
\begin{align}
  \mathcal{I}_n(\bm{\theta}) = \sum_{i=1}^n
  \tensor{\mathcal{S}_i(\bm{\theta})},
\end{align}
also used in the $BH^3$-algorithm.

Below we summarize the steps involved in calculating the score function

\begin{algorithmbis}[\emph{Calculating the score (\ref{eq:Scens})}]
\label{algo1}
For a single observation of responses $\bm{y}_i = (y_{i1},\ldots
y_{im})'$ and covariates $(\bm{x}_i',\bm{v}_i')'$, and
given a set of parameter values, $\bm{\theta}$:

\begin{enumerate}
\item a
\item b
\item c
\end{enumerate}
\begin{enumerate}
\item 
  Identify dichotomous responses and redefine failures as left censored at
  zero, and successes as right censored at zero.
\item Identify index of observed variables, $\mathcal{I}_i$ and
  censored variables $\mathcal{I}_i^C$, and let $\bm{y}_1 =
  (y_{ij})_{\{j\in\mathcal{I}_i\}}$ and $\bm{y}_0 =
  (y_{ij})_{\{j\in\mathcal{I}_i^C\}}$.  
\item Define matrix $\bm{L}$ as in (\ref{eq:Lmat}), i.e. the diagonal
  matrix with -1 at entry $i$, if the $i$th coordinate of
  $\bm{y_0}$ is right-censored, and ones on all other diagonal positions.
\item Calculate model-specific conditional mean $\mm$ and variance
  $\VV$ given covariates, and
  corresponding partial derivatives, and let
  $\bm{V}' = \frac{\partial\mvec\VV}{\partial\bm{\theta}'}$ and
  $\bm{m}' = \frac{\partial\mvec\mm}{\partial\bm{\theta}'}$.
\item For $(\bm{y}_1,\bm{y}_0)$ the
  underlying normal distribution of the fully observed process is
  partitioned as
  \begin{align}
    \mathcal{N}\left(
      \begin{pmatrix}
        \bm{m}_1 \\
        \bm{m}_0
      \end{pmatrix}, 
      \begin{pmatrix}
        \bm{V}_1 & \bm{V}_{01} \\
        \bm{V}_{10} & \bm{V}_1 
      \end{pmatrix}
    \right)
  \end{align}  
  Calculate the marginal means, $\bm{m}_1$, $\bm{m}_0$, and variance,
  $\bm{V}_1$, $\bm{V}_0$, of $\bm{y}_1$, by extracting
  elements $\mathcal{I}_i$. Similarly obtain matrix derivatives
  $\bm{V}'_1, \bm{V}'_0$ and $\bm{m}'_1, \bm{m}'_0$ by extracting
  the relevant rows from $\bm{V}'$ and $\bm{m}'$.
\item 
  Calculate $S_1(\bm{\theta})$, the score for $\bm{y}_1$, as in
  equation (\ref{eq:S}) based on $\bm{V}_1$ and $\bm{m}_1$
\item Calculate the
  conditional mean and variance of the underlying
  normal distribution of $\bm{y}_0$ given $\bm{y}_1$
  \begin{align}
    \widetilde{\bm{m}}_0 = \bm{m}_0 + \bm{V}_{01}\bm{V}_{0}^{-1}(\bm{y}_1-\bm{m}_1),
  \end{align}
  \begin{align}
    \widetilde{\bm{V}}_0 = \bm{V}_0 - \bm{V}_{01}\bm{V}_{0}^{-1}\bm{V}_{10},
  \end{align}
  and obtain the derivatives $\widetilde{\bm{m}}_0'$ and $\widetilde{\bm{V}}_0'$ as in (\ref{eq:dcondS})-(\ref{eq:dcondmu}).
\item Calculate
  \begin{align}
    S_2(\bm{\theta}) =
    \frac{1}{\Phi_{\bm{L}\widetilde{\bm{m}}_0,\bm{L}\widetilde{\bm{V}}_0\bm{L}}(\bm{L}\bm{y}_0)}
    \frac{\partial}{\partial\bm{\theta}'} 
    \cdff{\bm{L}\widetilde{\bm{m}}_0}{\bm{L}\widetilde{\bm{V}}_0\bm{L}}(\bm{L}\bm{y}_0)
  \end{align}
  using the results of Lemma (\ref{lem:1}) and (\ref{lem:2}) with
  derivatives 
  $$\frac{\partial}{\partial\bm{\theta}'}\bm{L}\widetilde{\bm{m}}_0 =
  \bm{L}\widetilde{\bm{m}}_0' \qquad\text{and}\qquad
  \frac{\partial}{\partial\bm{\theta}'} \bm{L}\widetilde{\bm{V}}_0\bm{L} = (\bm{L}\otimes\bm{L})\widetilde{\bm{V}}_0'.$$
\item Finally calculate the score contribution for the $i$th individual as
  $$\mathcal{S}(\bm{\theta};\bm{y}_i,\bm{x}_i,\bm{v}_i) = S_1(\bm{\theta}) + S_2(\bm{\theta}).$$
\end{enumerate}
\end{algorithmbis}



\section{Estimation via composite marginal likelihood}
\label{ref:complik}
The computational bottleneck of the MLE method is the calculation of
the derivatives of the normal CDF based on decomposing the
distribution function into all possible combinations of conditional
distributions. For the hessian this leads to a computational
complexity of order $\mathcal{O}(k^2)$. To remedy this problem in
models with large $k$, we propose to estimate model parameters via 
composite marginal likelihoods \citep{lindsay98:_compos}. 
We will compound over marginals defined by the index $\mathcal{K}$,
and define the composite marginal likelihood as
\begin{align}
  c\ell(\bm{\theta};\bm{y}) = \sum_{k\in \mathcal{K}} \ell_k(\bm{\theta}; \bm{y}^{(k)})
\end{align}
and the corresponding composite score
\begin{align}
  \mathcal{U}(\bm{\theta};\bm{y}) = \sum_{k\in \mathcal{K}} \mathcal{U}_k(\bm{\theta};\bm{y}^{(k)}),
\end{align}
where $\mathcal{U}_k$ is the score corresponding to the marginal
log-likelihood $\ell_k$.
We define
\begin{align}
  \mathcal{I}(\bm{\theta}) = \E(-\nabla\mathcal{U}(\bm{\theta};\bm{y}))
  \quad\text{and}\quad
\mathcal{J}(\bm{\theta}) = \var(\mathcal{U}(\bm{\theta};\bm{y})).
\end{align}
As we essentially have a mis-specified model, Bartlett's identity no
longer holds ($\mathcal{I}\neq \mathcal{J}$). Instead we can use the Godambe information
\begin{align}
  \mathcal{G}(\bm{\theta}) =
  \mathcal{I}(\bm{\theta})\mathcal{J}(\bm{\theta})^{-1}\mathcal{I}(\bm{\theta}),
\end{align}
and under general regularity conditions the composite likelihood
maximizer is asymptotically normally distributed 
$\sqrt{n}(\widehat{\bm{\theta}}_{cl}-\bm{\theta}^*)\overset{P}{\rightarrow}
\mathcal{N}(\bm{\theta}^*,G^{-1}(\bm{\theta}^*))$, with
$\bm{\theta}^*$ minimizing the Kullback-Leibler divergence to the
marginals of the true model distribution.

In principle the second derivative in $\mathcal{I}$ can be calculated
from the results of the previous sections, or it can be approximated
numerically.  Also Bartlett's identity holds for each marginal
likelihood term, i.e. $\mathcal{I}$ can be calculated as
\begin{align}
  \widehat{\mathcal{I}} = -\frac{1}{n}\sum_{i=1}^n\sum_{k\in\mathcal{K}}
  \mathcal{U}_k(\widehat{\bm{\theta}}_{cl};\bm{y}^{(k)}_i)^{\otimes 2}.
\end{align}
Similarly from the i.i.d. decomposition it follows that the empirical
estimator of $\mathcal{J}$ is
\begin{align}
  \widehat{\mathcal{J}} = \frac{1}{n}\sum_{i=1}^n
  \mathcal{U}(\widehat{\bm{\theta}}_{cl};\bm{y}_i)^{\otimes 2}.
\end{align}
In the composite likelihood approach we can perform the calculations
of the derivatives of the CDF for a fixed value $k_0$ corresponding to
the size of the composite blocks. Using all pairs the growth in the
data will be $k(k-1)/2$, and the asymptotic complexity is unchanged.
Selecting only neighboring pairs, e.g. $\{(1,2),(2,3),...\}$ will lead
to a $\mathcal{O}(n)$-algorithm. For consistency of the composite
likelihood estimator, all composite blocks must be correctly specified
and the composition must embrace enough of the parameter space in
order to identify the parameters of the full joint density. 

Generally the loss in efficiency might be compensated by a gain in
computational robustness and efficiency.  The proposed method of
compounding the likelihood of adjacent variables, has been analyzed in
a longitudinal setting \citep{joe09} showing reasonable power for an
auto-regressive correlation structure.  In essence this is a limited
information estimator but in contrast to \citep{muthen84limited}, we can in a
natural way include incomplete observations in the analysis and make
use of the increasing amount of inferential tools invented for composite
likelihoods (see \citep{varin11} for a recent overview).


\section{Usage}
\label{sec:usage}

The proposed method for estimation has been implemented in the
\proglang{R} package \pkg{lava.tobit} \citep{lavatobit} which acts as
a plug-in to the package \pkg{lava} \citep{holst12:_linear_laten_variab_model,lava} (an implementation of
the LLVM). 

\begin{figure}[ht]
  \centerline{
  \xymatrix{
    *+[F]{Y_1} & *+[F]{Y_2} & *+[F]{Y_3} \\ \\
    &
    *++[o][F-]{\eta_2}\ar[uul]^{\lambda_{1}=1}\ar[uu]^{\lambda_{2}}\ar[uur]_{\lambda_{3}}
    &
    \\ \\
    *+[F]{X_1}\ar[uur]^{\gamma_1}& &     *+[F]{X_2}\ar[uul]_{\gamma_2} \\
  }
}  
\caption{Path diagram of structural equation model defined by (\ref{eq:meas1}) and (\ref{eq:struct1}).}\label{eq:sem}
\end{figure}
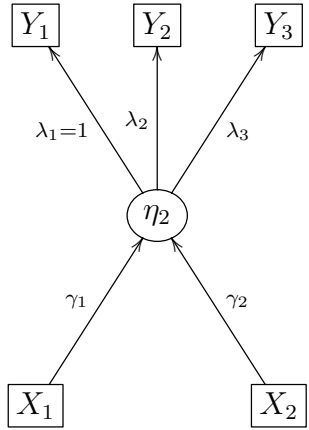

As an example we will define a simple structural equation model 
\begin{align}
Y_{ij} &=  \mu_j + \lambda_{j}\eta_i + \epsilon_{ij}\label{eq:meas1}, \\
\eta_i &= \gamma_1 X_1 + \gamma_2 X_2 + \zeta_i\label{eq:struct1}.
\end{align}
with $i=1,\ldots,n; j=1,\ldots,3$, and
$\epsilon_{ij}\sim\mathcal{N}(0,\sigma_j^2)$, $\zeta_i\sim\mathcal{N}(0,\xi^2)$.
The model is visualized in the path diagram in Figure \ref{eq:sem}.
Using the \pkg{lava}-model syntax, this model can be defined in
\proglang{R} as
\begin{singlespace}
\begin{Schunk}
  \begin{Sinput}
> library(lava.tobit)               
> m <- lvm()                        # Initialize 'lvm' model
> regression(m) <- c(Y1,Y2,Y3)~eta  # Measurement model
> latent(m) <- ~eta                 # Define 'eta' as random
> regression(m) <- eta~X1+X2        # Structural model  
\end{Sinput}
\end{Schunk}
\end{singlespace}
Various methods for adding (non-linear) parameter constraints and
covariance between residual terms of the model exists. We refer to
the \code{man}-pages of \pkg{lava} for further details.

To simulate 500 observations from the above model, where we dichotomize
$Y_2$ and introduce fixed right censoring time at 1.5 for $Y_3$, we
write
\begin{singlespace}
\begin{Schunk}
  \begin{Sinput}
> set.seed(1)
> d <- transform(sim(m,500),
+          Y2=factor(Y2>0),
+          Y3=Surv(ifelse(Y3<1.5,Y3,1.5),Y3<1.5))    
  \end{Sinput}
\end{Schunk}
\end{singlespace}
with intercepts set to 0 and all other parameters 1 (the default parameter
values of \code{sim}).  To find the MLE, the \code{estimate} function
is used, and since \code{Y2} and \code{Y3} were defined as a
\code{factor} (with two levels) and as a \code{Surv}-object, respectively,
the \code{estimate} method automatically calculates the likelihood and
score function for these parts (Probit and Tobit, respectively) using
the described method of Section 3.
\begin{singlespace}
\begin{Schunk}
  \begin{Sinput}
> estimate(m,d)
\end{Sinput}
\begin{Soutput}
                      Estimate Std. Error    Z value   Pr(>|z|)
Measurements:                                                  
   Y2<-eta           0.8695611  0.0913887  9.5149756     <1e-16
   Y3<-eta           1.0169726  0.0472970 21.5018283     <1e-16
Regressions:                                                   
   eta<-X1           1.0039318  0.0588643 17.0550265     <1e-16
   eta<-X2           1.0184532  0.0605893 16.8091257     <1e-16
Intercepts:                                                    
   Y2               -0.1033761  0.0869258 -1.1892456  0.2343430
   Y3               -0.0480167  0.0688087 -0.6978291  0.4852841
   eta              -0.0073254  0.0662993 -0.1104893  0.9120214
Residual Variances:                                            
   Y1                1.1241819  0.1155639  9.7277922           
   Y3                0.9106202  0.0987181  9.2244543           
   eta               1.1072682  0.1221184  9.0671677           
\end{Soutput}
\end{Schunk}
\end{singlespace}
To demonstrate the composite likelihood method, we dichotomize
\code{Y1}, and estimate the parameters of the model using the
\code{clprobit} function:
\begin{singlespace}
\begin{Schunk}
\begin{Sinput}
> d2 <- transform(d, Y1=factor(Y1>0))
> clprobit(m,k=2,data=d2)
\end{Sinput}
\begin{Soutput}
                       Estimate  Std. Error     Z value    Pr(>|z|)
Measurements:                                                      
   Y2<-eta           9.5979e-01  1.8883e-01  5.0828e+00  3.7186e-07
   Y3<-eta           9.2257e-01  1.4058e-01  6.5627e+00  5.2835e-11
Regressions:                                                       
   eta<-X1           9.9233e-01  1.4059e-01  7.0583e+00  1.6860e-12
   eta<-X2           9.1147e-01  3.1514e-01  2.8922e+00  3.8250e-03
Intercepts:                                                        
   Y2               -1.6560e-01  1.0419e-01 -1.5893e+00  1.1199e-01
   Y3                7.0581e-02  9.0536e-02  7.7959e-01  4.3563e-01
   eta              -1.2235e-01  9.5063e-02 -1.2870e+00  1.9808e-01
Residual Variances:                                                
   Y3                1.0816e+00  1.5267e-01  7.0847e+00            
   eta               1.0223e+00  1.6114e-01  6.3442e+00            
\end{Soutput}
\end{Schunk}
\end{singlespace}
In this example the block size was 2 and only adjacent variables were
compounded, e.g. $(Y_1,Y_2)$ and $(Y_2,Y_3)$. We refer to the
help-page of \code{clprobit} on how to customize the blocks used in
the composite likelihood estimation (as default continuous
non-censored variables will not be split up).




\section{Simulation study}
\label{sec:simulation}
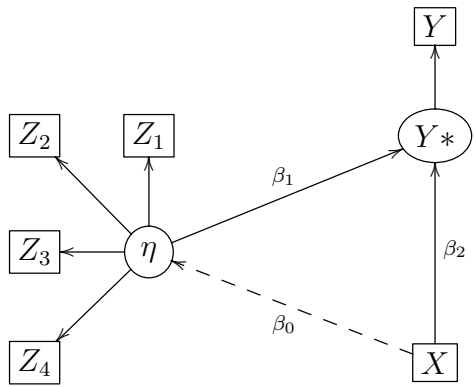
\begin{figure}[ht]
    \centerline{
    \xymatrix{
      & & & & *+[F]{Y} \\
      *+[F]{Z_2} & *+[F]{Z_1} & & & *++[o][F-]{Y*}\ar[u] \\
      *+[F]{Z_3} &
      *++[o][F-]{\eta}\ar[u]\ar[ul]\ar[l]\ar[dl]\ar[urrr]^{\beta_1}
      & & & \\
      *+[F]{Z_4} & & & & *+[F]{X}\ar[uu]_{\beta_2}\ar@{-->}[ulll]^{\beta_0} \\
    }
    }
    \caption{Path diagram of structural equation model defined by
      (\ref{eq:sim1}) and (\ref{eq:sim2}).}\label{fig:sim}
\end{figure}

In this section we will conduct a comparison between the limited
information estimator (LIE) proposed in \citep{muthen84limited} and MLE as obtained
by our method.
We will examine a model (see Figure \ref{fig:sim}) with a single binary
outcome, $Y$, which follows a Probit model given a latent variable
$\eta$ and a covariate $X$:
\begin{align}\label{eq:sim1}
  \Phi^{-1}\Big(\pr(Y=0\mid \eta, X)\Big) = \mu + \beta_1\eta + \beta_2X
\end{align}
The latent variable, $\eta$, is measured by 4 continuous variables
\begin{align}\label{eq:sim2}
  Z_j = \alpha_j + \lambda_j\eta + \epsilon_j, \qquad j=1,\ldots,4
\end{align}
and $\eta\sim\mathcal{N}(0,\xi^2)$,
$\epsilon_j\sim\mathcal{N}(0,\sigma_j^2)$ are pairwise independent.
We wish to estimate the effect of two covariates $\eta$ and $X$ on the 
binary outcome $Y$, which normally would be realizable within the 
Generalized Linear Model. However, we only have access to indirect
measurements of $\eta$, $Z_1-Z_4$, and using one of these variables as
predictor instead of $\eta$ will yield a biased estimate of the effect
of $\eta$, due to regression attenuation. The introduction of a
measurement model remedies this.

We simulated $n=500$ samples from the model with
$X\sim\mathcal{N}(0,1)$, $\lambda_j=\sigma_j^2=\xi^2=1$,
$\mu=\alpha_j=0$, $j=1,\ldots,4$, and the two parameters of primary
interest $\beta_1=1$ and $\beta_2=-0.5$.
\begin{table*}[ht]
  \centering
  \def\~{\hphantom{0}}
  \begin{minipage}{100mm}
  \caption{Simulation study comparing MLE and LIE
    \citep{muthen84limited} based on 10,000 replications of
    sample-size 500 from the model (\ref{fig:sim}) with 
    $\Phi^{-1}(\pr(Y=0\mid \eta, X)) = \mu + \beta_1\eta + \beta_2X$.
  } 
  \label{tab:beta2}
  \end{minipage}
  \begin{tabular}{llrrrrr}
    & & Variance & Rel. Eff. & Bias & MSE &
    $\frac{\text{Ave(SD}(\widehat{\beta}))}{SD(\widehat{\beta}_{sim})}$
    \\ \hline
    $\mu:$ \\ \hline
    &MLE &    0.0094 & 1    & 0.001 & 0.0094 & 1.00 \\
    &LIE A) & 0.0098 & 0.96 & 0.001 & 0.0098 & 1.03 \\
    &LIE B) & 0.0099 & 0.95 & 0.001 & 0.0099 & 1.02 \\
    &LIE C) & 0.0100 & 0.94 & 0.001 & 0.0100 & 1.03 \\
    &LIE D) & 0.0098 & 0.96 & 0.001 & 0.0098 & 1.03 \\
    \hline
    $\beta_1:$ \\ \hline
    &MLE &    0.0128 & 1    & 0.020 & 0.0132 & 1.00 \\
    &LIE A) & 0.0139 & 0.93 & 0.023 & 0.0144 & 1.03 \\
    &LIE B) & 0.0146 & 0.88 & 0.025 & 0.0152 & 1.02 \\
    &LIE C) & 0.0188 & 0.68 & 0.030 & 0.0197 & 1.03 \\
    &LIE D) & 0.0139 & 0.93 & 0.023 & 0.0144 & 1.03 \\
    \hline
    $\beta_2:$ \\ \hline 
    &MLE &    0.0077 & 1    & -0.009 & 0.0078 & 0.99 \\
    &LIE A) & 0.0123 & 0.63 & -0.009 & 0.0123 & 1.02 \\
    &LIE B) & 0.0087 & 0.89 & -0.011 & 0.0088 & 1.01 \\
    &LIE C) & 0.0163 & 0.47 & -0.015 & 0.0165 & 1.03 \\
    &LIE D) & 0.0083 & 0.93 & -0.010 & 0.0084 & 1.01 \\ 
    \hline
  \end{tabular}
\end{table*}
The MLE and LIE were compared for model
(\ref{eq:sim1})-(\ref{eq:sim2}) (model A). We also calculated the LIE
for the model (model D) where a direct effect of $X$ on $\eta$ was
included (the parameter $\beta_0$ in Figure \ref{fig:sim}, a model
where a covariance parameter between $\eta$ and $x$ was included
(model B), and a model where covariance between the $\eta$ and $x$ was
included but fixed to zero (model C). The results of the simulation
study based on 10,000 replications are shown in Table \ref{tab:beta2}.

As the MLE asymptotically defines the Cramer-Rao lower-bound, we
calculated the relative efficiency as the ratio between the variance
of the Monte Carlo parameter estimates of the MLE and the variance of
the LIE estimates. For the intercept parameter, $\mu$, all estimators
performed well with practically no bias and a relative efficiency of
the LIE around 0.95.  The LIE of model A and D had a comparable
performance with a relative efficiency of 0.93 for the parameter
$\beta_1$. Remarkably the LIE of model C performed significantly worse,
even though it should be equivalent with model A. This might be an
implementation artifact.  With a sample-size of $n=500$ we still saw a
little bias for the parameter $\beta_1$ (smallest for the MLE).
For $\beta_2$ the bias of all estimators was acceptable, but to our
surprise the LIE for the true model (A) performed much worse than model D
(relative efficiency 0.93) and B (relative efficiency 0.89).

The performance of the estimates of the standard errors of the
parameters was quantified by calculating the ratio between the average
standard error estimate and the standard deviation of the Monte Carlo
parameter estimates. The standard error estimates generally performed
acceptable for both the LIE and MLE (calculated via the outer product
of the score). Perhaps there is a tendency towards the standard errors
from the LIE being a little too conservative.

Overall we can conclude that model D, where the structural model was
modeled as $\eta = \beta_0X + \zeta$,
yielded the best results for the LIE even under the model
$\beta_0=0$. While clearly outperformed by the MLE, the loss in
efficiency might for some applications be out-weighted by a gain in
computational efficiency of the LIE. On the other hand, the lack of access
to likelihood ratio testing, profile likelihood confidence intervals,
the ability to handle incomplete observations, etc., might render this
option less attractive.

All simulations were conducted in \proglang{Mplus} version 5.1 \citep{mplus5} and
\pkg{lava.tobit} version 0.4-7 \citep{lavatobit}.



\section{Application}
\label{sec:application}
In this section we will demonstrate the modeling framework on brain
imaging and personality data. Understanding the relationship between
personality and biological processes in the brain is recognized as
being an important step towards understanding the etiology of mental
disorders such as schizophrenia and depression.  Dysfunction of the
serotonin (5-hydroxy\-tryp\-t\-a\-mine (5-HT)) \hta receptors is
considered involved in the development of such disorders
\citep{williams96:_assoc_t102c,choi04:_assoc_g}, thus motivating the
search for personality traits associated with the serotonergic system.

Cloninger \citep{cloninger87} developed the \emph{Temperament and
  Character Inventory} (TCI) personality model with a biological
interpretation derived from animal studies. The TCI scale is based on
dichotomous questionnaires which are summarized into personality
traits describing different patterns of behavior such as \emph{Harm
  Avoidance}, \emph{Reward Dependence} and \emph{Novelty seeking}. It
was originally hypothesized that the neurotransmitter systems were
related uniquely to specific traits, for instance \emph{Harm
  avoidance} and the serotonergic system. However, it has later been
proposed that the serotonergic system could be related to regulation
of other personality traits \citep{paris05:_neurob} as suggested by
both genetic and imaging studies \citep{ebstein97:_ht2c_htr2c,
  Goethals2007455}. The trait Reward Dependence ($RD$) characterizes
how a person reacts to reward or punishment. It also seems that
serotonin plays an active role for how people reacts towards
reward. For example, animal studies have shown that decreased
serotonin levels cause more frequent impulsive choices
\citep{pmid10550490,pmid10823413}.

In this study we therefore wanted to examine the relationship between
measurements of the serotonergic system in the human brain and the
trait Reward Dependence. A Danish translation of the self-administered
TCI-R questionnaire \citep{cloninger94:tci} was answered by 170
subjects. Reward Dependence is traditionally quantified as the sum of
three sub-dimensions \emph{Sentimentality} ($RD_1$, 10 questions),
\emph{Attachment} ($RD_3$, 8 questions), and \emph{Dependence}
($RD_4$, 6 questions). Each of the sub-dimensions are calculated as
the sum of the related dichotomous questions.  As a preliminary
exercise, we examined the underlying factor structure that justifies
the use of these summary statistics.  The requirements of criterion
related construct validity \citep{rosenbaum89} are
\emph{uni-dimensionality}, \emph{monotonicity} and \emph{local
  independence} (conditional independence given the latent variable)
as illustrated in (\ref{eq:fac1})
\begin{align}\label{eq:fac1}
  \xymatrix{
    *+[F]{Y_1} & *+[F]{Y_2} & \cdots & *+[F]{Y_p} \\
     *++[o][F-]{Y_1^*}\ar[u] & *++[o][F-]{Y_2^*}\ar[u] &
     \cdots\ar@{-->}[u] &  *++[o][F-]{Y_p^*}\ar[u] \\ \\
     & *+++=[o][F-]{{~}\eta_{~}}
    \ar[uul]^{\lambda_{1}}\ar[uu]^{\lambda_{2}}\ar@{-->}[uur]\ar[uurr]_{\lambda_{p}}
  }
\end{align}
with
\begin{align}
  \Phi^{-1}\Big(\pr(Y_i=0|\eta)\Big) = \mu_i + \lambda_i\eta.
\end{align}
Evidence against this model structure was tested with a likelihood
ratio test against the saturated model with a free mean and
correlation structure.  The factor analysis of $RD_1$ and $RD_3$
indicated some problems with these scales and we therefore focused on
$RD_4$.  One question (number 71) had to be removed from the analysis
due to too little variation causing numerical instability.
Three individuals had incomplete observations, which was handled in
the MLE approach by ignoring the missing data mechanism under a
missing at random assumption \citep{MR1925014}. The $\chi^2$
omnibus-test yielded a p-value of $0.58$ thus not showing any evidence
against the suggested factor structure. For a reliable scale, we would
further expect all factor loadings to be equal (after appropriate
coding of the questions). A likelihood ratio test against this model
yielded a p-value of $0.61$. This suggests that $RD_4$ is a
valid scale. The dimension \emph{Dependence} has also previously been
shown to be a valid and reliable scale in a sample of hospitalized and
ambulant Danish psychiatric patients \citep{thorleifsson10:_person}. In the
subsequent analyses, we therefore used the suggested factor structure
for $RD_4$ with equal factor loadings.

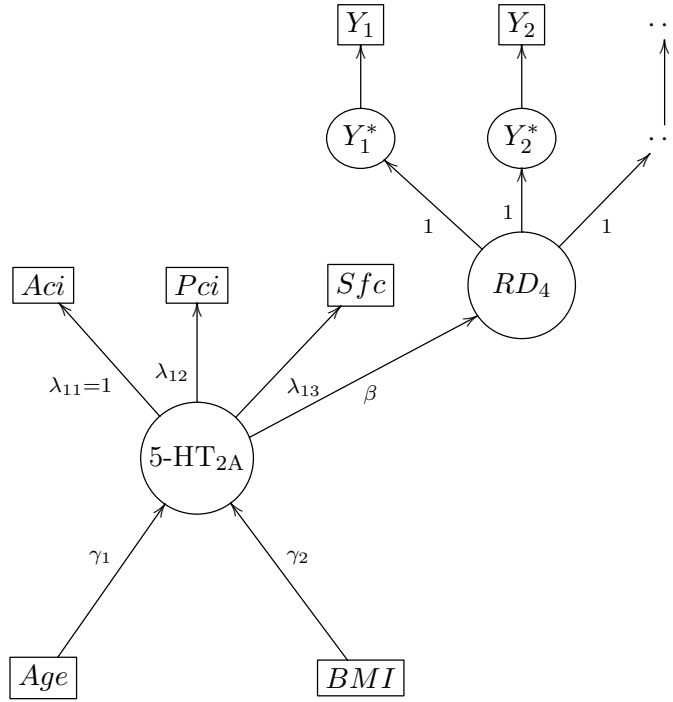
\begin{figure}[ht]
  \small
  \centerline{
    \xymatrix {
      & & *+[F]{Y_1} & *+[F]{Y_2} & \cdots  \\
      & & *++[o][F-]{Y_1^*}\ar[u] & *++[o][F-]{Y_2^*}\ar[u] &
      \cdots\ar[u] \\
      *+[F]{Aci} & *+[F]{Pci} & *+[F]{Sfc} & *++++=[o][F-]{RD_4}
      \ar[ul]^{1}\ar[u]^{1}\ar[ur]_{1}
      \\
      &
      *++=[o][F]{\hta}\ar[ul]^{\lambda_{11}=1}\ar[u]^{\lambda_{12}}\ar[ur]_{\lambda_{13}}\ar@<-0.5ex>[urr]_{\beta} \\
      \\
      *+[F]{Age}\ar[uur]^{\gamma_1} & & *+[F]{BMI}\ar[uul]_{\gamma_2}
    }
  }
  \caption{Structural equation model describing the association
    between \hta receptor binding potential and Reward Dependence.}
  \label{eq:tci2a}
\end{figure}

Our aim was to quantify the association between central 5-HT function
and the trait \emph{Dependence}. Unfortunately it is not possible to
measure brain 5-HT levels directly \emph{in vivo}. However, several studies
have demonstrated a (negative) linear relationship between brain 5-HT
levels and 5-HT2A receptor binding \citep{pmid18657534,pmid17124620},
suggesting that low 5-HT leads to compensatory up-regulation of 5-HT2A
receptor binding. We therefore proposed a measurement model where \hta
receptor binding potential measurements were perceived as indirect measurements
of underlying global 5-HT level:
\begin{align}
  -{\hta}_{,j} = \mu_j + \lambda_j\cdot\text{5-HT} + \epsilon_j.
\end{align}
\hta receptor binding potential potential (\BPp) was quantified by Positron
Emission Tomography (PET) techniques and segmented into summary
statistics of anatomically defined regions
\citep{svarer05}. For our analysis we chose three regions of interest
which previously have been shown to yield reliable quantification of
\hta receptor binding: Superior frontal cortex (sfc), Anterior
cingulate gyrus (acg), and Posterior cingulate gyrus (pcg) (see Figure
\ref{eq:tci2a}). Effects of age and BMI have previously been demonstrated
and was therefore added to the structural model (with BMI dichotomized
as overweight (BMI>25) vs non-overweight):
\begin{align}
  \text{5-HT} \propto \hta = \gamma_1\cdot\text{Age} + \gamma_2\cdot\text{BMI} + \zeta.
\end{align}
We conducted an analysis of 56 subjects who completed the
TCI questionnaire and also underwent PET examination of the \hta
receptor binding potential. We refer to the paper \citep{DavidErritzoe03032010} for
a description of this PET sample and details on the methodology.
The measurement model for the \hta receptor measurements was combined
with the measurement model of \emph{Dependence} (see
(\ref{eq:fac1})). The two measurement models were
linked with a regression between the two latent variables
\begin{align}
  \E(RD_4\mid\hta) = \beta\cdot\hta,
\end{align}
leading to the model described in the path-diagram of Figure \ref{eq:tci2a}.

  The parameters of the model were estimated with MLE as described in
Section \ref{sec:method}. In the analysis we had to remove one
additional item from the measurement model of $RD_4$ (question number
131) to avoid numerical instability.

The regression coefficient $\beta$ is interpreted as the linear effect
of the underlying \hta receptor binding potential level on the latent personality trait
\emph{Dependence}. This interpretation becomes clearer using the
standardized coefficient, which we estimated to a increase of 0.791
standard deviation in \emph{Dependence} caused by increasing the \hta
receptor binding potential
levels one standard deviation (p-value 0.04). Hence high 
\hta receptor binding levels (low 5-HT levels) are associated with
higher values of \emph{Dependence} (more sensitive and socially dependent
individuals).

\begin{figure}[htbp]
    \centering
    \includegraphics[height=7cm,keepaspectratio=true]{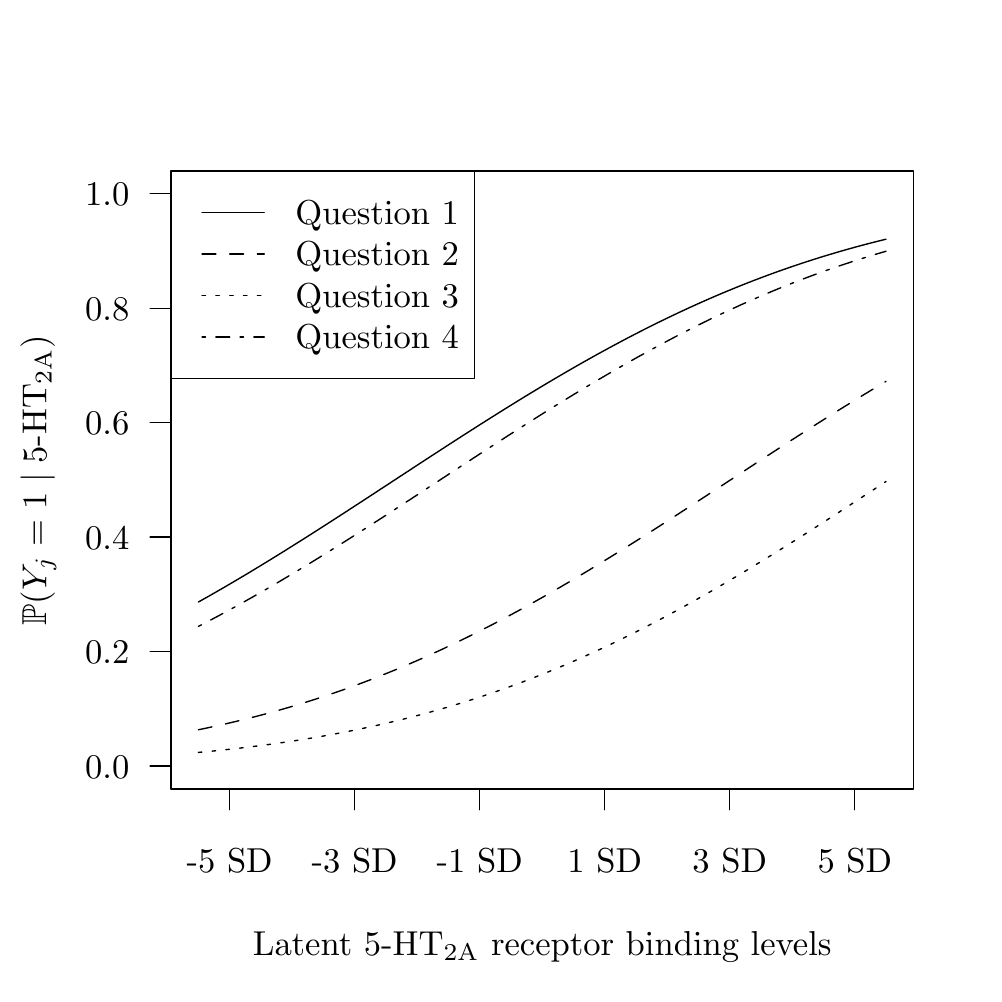}
    \caption{Probabilities of answering yes to different questions of
      the \emph{Dependence} personality trait, with varying levels of
      latent 5-HT levels as measured by \hta receptor binding
      potential. The x-axis is measured in standard deviations away
      from the mean (reference: non-overweight, age 36).}
  \label{fig:probplot}
\end{figure}

The model also allows us estimate the conditional
probabilities of the answers to the different questions given the 5-HT
levels. For instance, for the first question in the $RD_4$, we have
\begin{align*}
  \pr(Y_1 = 1\mid \hta) = \Phi(0.42+1\cdot 0.42\hta).
\end{align*}
See Figure \ref{fig:probplot}.  Similarly the joint probability of
different combinations of questions could be evaluated. Also, to
obtain a better understanding of the 5-HT/\hta receptor 
effect, we can calculate the probability ratio of answering positively
to the first question, for a subject with average \hta receptor levels and a
subject with \hta receptor levels two standard deviations away from the average,
$2\cdot \text{SD}(\hta\mid Age,BMI)=0.96$,
\begin{align*}
  \frac{\pr(Y_1 = 0\mid \text{5-HT}=0.96)}{\pr(Y_1=0\mid
    \text{5-HT}=0)} = 1.20
\end{align*}
indicating a 20\% higher probability of answering positively to question
1 for the subject with high levels of \hta receptor binding potential
(95\% confidence limits by the Delta method: $[1.05; 1.34]$). 



\section{Conclusion}
\label{sec:conclusion}
Various methods for estimating parameters in models with correlated
multivariate binary data has been proposed.  Monte Carlo based methods
such as the MCEM algorithm and Bayesian methods are possible, but
convergence can be slow and monitoring of the convergence of the
Markov chains to the stationary distribution and assessment of the
influence of prior distributions in the Bayesian context remains
controversial topics.  Deterministic integration such as AGQ also
suffers from ad hoc decisions which must be made regarding the
number of quadrature points to achieve a reasonable approximation. For
complex models with many latent variables, it is well-known that this
approach may break down in practice.

Our method restricts itself to models with a Probit link, and we have
demonstrated how this leads to an algorithm, which allows us to
perform fast and precise evaluations of the score and likelihood
function of the model. Implementation of generalizations to estimators
based on score equations such as Inverse Probability Weighting should
therefore follow immediately. In contrast for methods based on
applying the Laplace approximation or AGQ where numerical derivatives
of the integrated log-likelihood typically is used to approximate the
score function, such generalizations could prove much more
difficult. Our proposal is most well-suited for models with a moderate
number of observed items and possible very complex latent
structure. However, the composite likelihood approach scales well with
dimension in both number of items and latent variables, and may serve
as an important framework for the estimation of complex structural
equation models with non-normal response variables.

A strong feature of our model is the possibility of combining
continuous, dichotomous and censored observations in a simultaneous
model. This may serve as an important framework for the examination of
causality \citep{ditlevsen05:_mediat_propor}.

The model has been implemented in full generality in the \pkg{lava} and
\pkg{lava.tobit} packages \citep{lavatobit} available freely on
\url{http://r-forge.r-project.org}.



\section{Acknowledgments}
This work was supported by The Danish Agency for Science, Technology
and Innovation.

\addcontentsline{toc}{section}{Bibliography}
\bibliographystyle{elsarticle-harv}
\bibliography{all}

\end{document}